# Hydrogen storage in Li functionalized [2,2,2]paracyclophane at cryogenic to room temperatures: A computational quest

Rakesh K. Sahoo, Sridhar Sahu

Computational Materials Research Lab, Department of Physics, Indian Institute of Technology (Indian School of Mines) Dhanbad, India

## Abstract

In this work, we have studied the hydrogen adsorption-desorption properties, and storage capacities of Li functionalized [2,2,2]paracyclophane (PCP222) using dispersion-corrected density functional theory and molecular dynamic simulation. The Li atom was found bonded strongly with the benzene ring of PCP222 via Dewar interaction. Subsequently, the calculation of the diffusion energy barrier revealed a significantly high energy barrier of 1.38 eV, preventing the Li clustering on PCP222. The host material, PCP222-3Li adsorbed up to 15$H_2$ molecules via charge polarization mechanism with an average adsorption energy of 0.145 eV/5$H_2$, suggesting physisorption type of adsorption. The PCP222 functionalized with three Li atom showed maximum hydrogen uptake capacity up to 8.32 wt% which was fairly above the US-DOE criterion. The practical $H_2$ storage estimation revealed that the PCP222-3Li desorbed 100% of adsorbed $H_2$ molecules at the temperature range of 260 K-300 K and pressure range of 1-10 bar. The maximum $H_2$ desorption temperature estimated by the Vant-Hoff relation was found to be 219 K and 266 K at 1 bar and 5 bar, respectively. The ADMP molecular dynamics simulations assured the reversibility of adsorbed $H_2$ and the structural integrity of the host material at sufficiently above the desorption temperature (300K and 500K). Therefore, the Li-functionalized PCP222 can be considered as a thermodynamically viable and potentially reversible $H_2$ storage material below room temperature.

**Keywords:** Hydrogen storage, DFT, Van't-Hoff equation, ADMP, [2,2,2]paracyclophane, PCP222, ESP

## 1 Introduction

The excessive consumption of traditional fossil fuels has not only led to the depletion of the energy supplies but also has emerged as the prime cause of environmental pollution. The global consumption of petroleum and other traditional fossil fuel is anticipated to expand up to 56% by the year 2040 and the crude oil supply is expected to endure until 2060 if the current demand trend continues[1]. Thus, it is essential to develop alternative energy sources that are free from the drawbacks of traditional fossil fuels. To meet the world's energy demand and reduce the pollution caused by fossil fuels, hydrogen has been considered as a plausible alternative due to its natural abundance, environmental friendliness, and regenerative properties. One of the distinctive quality of hydrogen is that it produces a large amount of energy per unit mass (120 MJ/kg) without releasing any pollutant by-products [2, 3]. Despite these benefits, however, the use of hydrogen in practice is limited due to the obstacle of finding the most appropriate and affordable way to store and deliver hydrogen under normal environmental conditions. As per the criteria proposed by the United State department of energy (DOE-US) an effective hydrogen storage material should have a minimum storage capacity of up to 5.5 wt% by the year 2025 at moderate thermodynamics[5, 6]. In addition, as reported by many authors, the adsorption energy of hydrogen molecules of an effective storage materials should be in the range of 0.1 eV/$H_2$ to 0.6 eV/$H_2$[4].

Though, numerous varieties of materials such as; metal hydrides [7, 8], graphene [9, 10], metal alloys [11, 12], metal-organic frameworks (MOF) [13, 14], covalent-organic frameworks (COF) [15] and carbon nanostructures [16, 17] etc have been investigated both theoretically and experimentally as potential hydrogen storage materials, but there are many drawbacks and unsolved issues to handle. The metal hydrides and complex hydrides store hydrogen via chemisorption process which is highly irreversible and prevents easy desorption of hydrogen [18]. For example, Al(BH4)3, which yields hydrogen uptake capacity of up to 16.9 wt%, has high desorption temperature (about 1000 K) that makes the material non-effective practical reversible hydrogen storage applications[19]. Under ambient conditions, Mg-metal hydrides have a storage capacity up to 7.6 wt%; however, it can only be used for 2-3 cycles[20]. Tavhare *et al.* studied the hetero atom substituted Ti-benzene and reported an $H_2$ uptake capacity up to 5.85 wt%, but at relatively high desorption temperature (1193 K)[21]. Furthermore, MOF and COF applications are constrained in the practical $H_2$ storage field due to the difficulties of their heavy structure and challenging step-wise production [22].

The efficient use of carbonaceous materials as hydrogen storage media was initially reported by Dillon *et al.* [23]. Carbonaceous materials are appropriate for $H_2$ storage due to their unique qualities such as, large surface area, high porosity, better stabilities, and low densities. However, the early findings have shown that these pure materials are weakly interact with the hydrogen molecules (with BE ~4-5 kJ/mol), thus impractical for realistic hydrogen storage at ambient environment [24, 25]. Meanwhile, carbon-based pure substrates are excellent materials for hydrogen storage at cryogenic temperatures. For instance, pure single wall carbon nanotube

(SWCNT) can store hydrogen molecules up to 8.25 wt%, with a substantially lower desorption temperature of 80 K [26].

It has been reported that the $H_2$ interaction strength and the desorption temperature can be tuned by integrating pure carbon substrates with alkali metal (AM)(Li, Na, and K), alkali earth metals (Be, Mg, Ca), and transition metals (TM)(Sc, Ti, V, Y.)[27, 28, 29]. Numerous theoretical investigations showed that integrating AMs and TMs with the carbon/borane substrates can bind $H_2$ molecules via charge polarization and the Kubas mechanism [30, 31]. The metallic atom decorated fullerenes were first explored to investigate the impact of metal integration on pure carbon substrates. According to studies by Sun *et al.*, Li decorated fullerene could show a storage capacity of 9 wt%; however, the hydrogen adsorption energy was estimated to be 0.075 eV/$H_2$, which is much lower than the DOE criterion [32]. The Li and Na-loaded $C_{60}$ revealed $H_2$ uptake capacities of 4.5 wt% and 4 wt%, respectively, that were significantly below the target of DoE [33]. Experimental studies of transition metals like V and Pd decorated CNT reveal 0.66 wt% and 0.69 wt% of hydrogen capacity respectively, while pure CNT has 0.53 wt% of storage capacity [34]. Sahoo *et al.* reported storage of $H_2$ on Li and Sc doped $C_8N_8$ cage via Niu-Rao-Jena and Kubas interaction and estimated a desorption temperature of 286 K and 456 K, respectively [35]. The Li and Na decorated on $C_{24}$ fullerene could adsorb $H_2$ molecules, with average hydrogen binding energies of 0.198 eV/$H_2$ and 0.164 eV/$H_2$ and led to storage capacity up to 12.7 wt% and 10 wt %, respectively [36]. Recently, we have investigated the $H_2$ storage on alkali metal decorated $C_{20}$ fullerene and found the molecular hydrogen are physisorbed on host material via charge polarization mechanism with desorption temperature of 182 -191 K [37]. Each Li and Na atom on $C_{20}$ could uptake up to 5$H_2$ molecules with a total gravimetric storage capacity of 13.08 wt % and 10.82 wt%, respectively, and the $H_2$ binding energies found in the range of 0.12 eV—0.13 eV/$H_2$.

Other carbonaceous materials such as functionalized organometallic compounds, macrocyclic compounds have also been reported recently as potential candidates for hydrogen storage. For example, Mahamiya *et al.* revealed the $H_2$ storage capacities of 11. 9 wt % in K and Ca decorated biphenylene with an average adsorption energy of 0.24-0.33 eV [38]. Y atom doped zeolite shows high capacity adsorption of $H_2$ with binding energy 0.35 eV/$H_2$ and the desorption energy of 437K for fuel cells[39]. Lithium-doped Calixarenes show an excellent hydrogen storage behaviour but at very low up to 100 K [40]. Calix[4]arene functionalized with Li metal reveals 10 wt% storage capacity via Kubas—Niu—Rao—Jena interaction, and all most all $H_2$ desorbed at a temperature of 273 K [41].

Macrocyclic compounds such as, paracyclophane (PCP), a subgroup derivative of cyclophanes, contains aromatic benzene rings, and their nomenclature is established on the arene substitution pattern. For a [n,n]paracyclophane, the number of -$CH_2$- moiety connecting the successive benzene rings is indicated by the number in the square bracket [42]. Due to the existence of aromatic benzene rings in the geometry, PCPs are easy to synthesize experimentally and can be functionalized with metal atoms, making them a viable choice for hydrogen storage candidates. A report on Li and Sc functionalized [4,4]paracyclophane revealed the hydrogen uptake capacity up to 11.8 wt% and 13.7 wt% with an average adsorption energy of 0.08 eV/$H_2$ and 0.3 eV/$H_2$ respectively [43]. Sahoo *et al.* recently studied the $H_2$ storage capacity of [1,1]paracyclophane functionalized with Sc and Y metals and found an $H_2$ gravimetric storage capacity of 8.22 wt% and 6.33 wt%, respectively, with an average adsorption energy 0.36 eV/$H_2$[44]. They reported the $H_2$ desorption temperature of 439 K and 412 K for Sc and Y doped PCP11, respectively, at 1 atm. The hydrogen molecules are physisorbed on Li, and Sc decorated paracyclophane via Kubas-Niu-Jena interaction and show a storage capacity of 10.3 wt%, as reported by Sathe *et al.* [45]. Many more alkali metal-doped macrocyclic compounds have also been investigated for hydrogen storage candidates and found the storage capacity above the DOE target; however very few reported the practical $H_2$ capacity at various thermodynamic conditions[46, 47]. Though few of PCP-based hydrogen storage systems are available in literature, the [2,2,2]paracyclophane (PCP222) which is experimentally synthesized by Tabushi *et al.*[48] is yet to be explored as hydrogen storage material. Because Li, the lightest alkali metal atom and can hold $H_2$ molecules via charge polarization mechanism, it can serve as better sorption center on PCP.

Therefore, in the current work, we intend to investigate the hydrogen storage properties and potential of Li functionalized [2,2,2]paracyclophane (PCP222). We chose the PCP222 for hydrogen storage because it is already experimentally synthesized and can be decorated with metal atoms to form a hydrogen storage media with a high hydrogen uptake capacity. The Li atoms are functionalized as sorption centers; this is because the light-weight metal doping method is an effective way to increase the capacity of $H_2$ storage. Li being the lightest alkali metal atom, received a lot of attention to for hydrogen sorption application. Though there are few reports available based on hydrogen adsorption mechanism on metal doped macrocyclic organic molecules and other Li decorated nanostructures, our work is the first to reveal the efficiency of Li functionalized PCP222 using the atomistic MD simulation, practical storage capacity and diffusion energy barrier estimation

## 2 Theory and Computation

The theoretical computations are carried out on [2.2.2] paracyclophane (PCP222) and their hydrogenated derivatives within the framework of density functional theory (DFT)[49]. The modern range separated hybrid functional *wB97Xd* is used, and molecular orbitals (MO) are defined as linear combination of atom centered basis functions, with all atoms using the valence diffuse and polarization function 6-311+G(d,p) basis sets. The *wB97Xd*, a long range separated form of Becke's 97 functional, also adds Grimme's D2 dispersion correction[50, 51]. It is worth mentioning that the *wB97Xd* is a reliable approach to investigate the non-covalent interaction of metal doped organic molecules and their thermochemistry. The harmonic frequencies of all the studied structures are calculated to confirm that they are truly in the ground state on the potential surface.

Some of the crucial quantitative metrics, including, binding energy of metal atom on host, average $H_2$ adsorption energy and successive $H_2$ desorption energy must be determined in order to analyze the mechanism of hydrogen storage.

The binding strength of Li atom on the PCP222 is calculated by the following expression[44];

$$E_b = \frac{1}{m}[E_{PCP222} + mE_{Li} - E_{PCP222+mLi}] \qquad (1)$$

Where $E_{PCP222}$, $E_{Li}$, and $E_{PCP222+mLi}$ are symbolize for the total energy of PCP222, energy of single isolated Li atom and energy of Li-decorated PCP222 respectively. *m* denotes for the number of Li atoms used to functionalized the PCP222.

The average adsorption energy of $H_2$ molecules with Li functionalized PCP222 is estimated as [52];

$$E_{ads} = \frac{1}{n}[E_{PCP222+mLi} + nE_{H_2} - E_{PCP222+mLi+nH_2}] \qquad (2)$$

Where, $E_{H2}$, and $E_{PCP222+mLi+nH2}$ represents the energy of isolated single $H_2$ molecule and hydrogen adsorbed $PCP222+mLi$, respectively. $n$ is the number of $H_2$ molecules adsorbed in each Li functionalized PCP222.

The successive desorption energy of adsorbed $H_2$ molecules is estimated using following equation[52].

$$E_{des} = \frac{1}{n}\left[E_{H_2} + E_{Host+(n-1)H_2} - E_{Host+nH_2}\right] \quad (3)$$

where $E_{Host+(n-1)H_2}$ is the energy of previous $H_2$ molecules adsorbed $E_{Host+nH_2}$.

The energy gap between the highest occupied molecular orbital (HOMO) and the lowest unoccupied molecular orbital (LUMO) is calculated to ensure the kinetic stability of the Li functionalized PCP222 and their hydrogen derivatives. The Hirshfeld charges and electrostatic potential map (ESP) was used to study electronic charge transfer and interaction mechanism. Further, to understand the metal and hydrogen interaction we have performed the partial density of states (PDOS), and topological using the Bader's quantum theory of atoms in molecules (QTAIM). To investigate the structural integrity of the host material and H2 reversibility of the system, atomistic molecular dynamic simulations were carried out using the expanded lagrangian approach, atom-centered density matrix propagation (ADMP).

To determine the H2 adsorption capacity, gravimetric density (wt%) of hydrogen can be calculated using the following expression [53]:

$$H_2(wt\%) = \frac{M_{H_2}}{M_{H_2}+M_{Host}} \times 100 \quad (4)$$

Here $M_{H2}$ represent the mass of the total number of $H_2$ molecules adsorbed and $M_{Host}$ represent the mass of Li functionalized PCP222.

## 3 Results and Discussion

### 3.1 Structural properties of PCP222

Figure 1 depicts the ground state geometrical structure of PCP222. The PCP222 comprises three benzene rings, that are linked via two $CH_2$ moiety as bridge between the adjacent rings. The lengths of the nearest $CH_2$-$CH_2$, and the $CH_2$ across the benzene rings are observed to be 1.5 and 5.84 Å, respectively, that agrees with the empirically reported value by Cohen-Addad *et al* [54]. To confirm the aromaticity of the relaxed PCP222, we calculated the Nucleus Independent Chemical Shift (NICS) from center to to 3 Å above the benzene ring by increment of 1 Å. The NICS(1) is found to have negative maximum (-10.1 ppm), demonstrating the aromatic character of PCP222[55, 56]. This suggest that the cyclic rings of PCP222 are $\pi$-electron rich and most probably can bind the metal atom above (outside of PC222) the benzene rings. The Li atom then functionalized above the benzene rings and on every possible site of PCP222 and allowed to relax as discussed below.

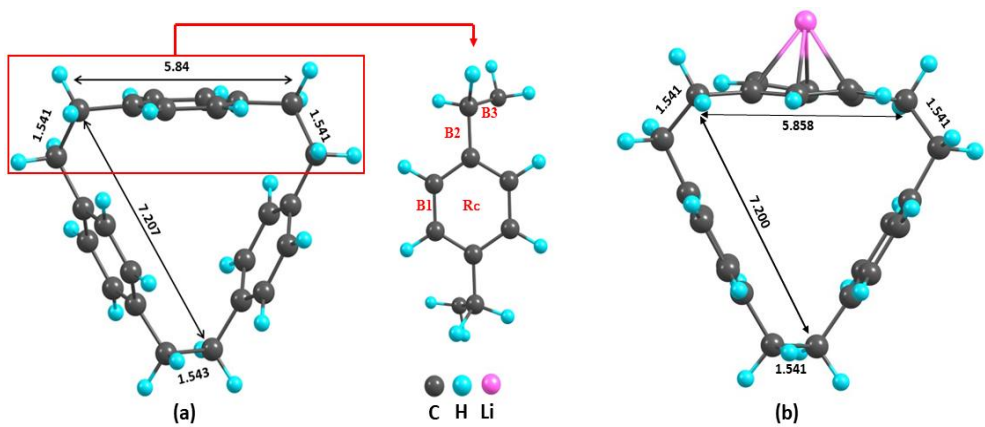

**Figure 1:** (a) Optimized structure of PCP222 with adsorption site marked with red-colored text, (b) Li functionalized PCP222.

### 3.2 Functionalization of Li atom on PCP222

To explore the hydrogen adsorption capacity in Li-functionalized PCP222, we must first carefully examine the suitable adsorption site for Li atoms on the PCP222. In order to do this, we investigated several PCP222 adsorption site, including the C-C bridge of benzene ring (B1), $CH_2$ moiety and benzene bridge (B2), $CH_2$ - $CH_2$ bridge (B3), and above the center of benzene ($R_c$). All the possible Li adsorption sites of PCP222 are depicted in Figure 1(a). A single Li atom is placed nearly 2 Å above the several probable adsorption sites of PCP222 and the structure is allowed to get optimized. It is observed that functionalization of Li atom over B1 and B2 sites, it migrate towards the $R_c$ site following the optimization. On optimization of Li atom over B3 site, the it moves away from the PCP222 and does not bind to the surface. We found that the Li atom is stable on $R_c$ site with binding energy of 0.32 eV that is 0.1 eV higher than that of Li on PCP44, reported by Sathe *et al.* [43]. The Li atom is supposed to be functionalized on PCP222 via Dewar mechanism, in which is due to the electronic charge transfer between the *p*-complex and s- orbitals of Li atom [43, 45]. After functionalization of Li, the estimated Hirshfeld charge on benzene ring of PCP222 is increased to -0.08 e.u from -0.03 e.u (in bare PCP222). These charges are transferred from the metal atom, with the Hirshfeld charges on Li atom being +0.35 e.u after functionalization, which make the Li atom ionic. The ionic Li atom is exposed to the guest $H_2$ molecules and can bind them via charge polarization mechanism as proposed by Niu *et al.* [57]. No significant change in geometrical bond distances is observed after the functionalization of Li. The thermal stability of the structures (host) is discussed in the molecular dynamic simulations section (section 3.5). All the hydrogen adsorption/desorption simulations are performed by functionalizing the Li atom above the center of benzene ring of PCP222.

### 3.2.1 Diffusion energy barrier calculation

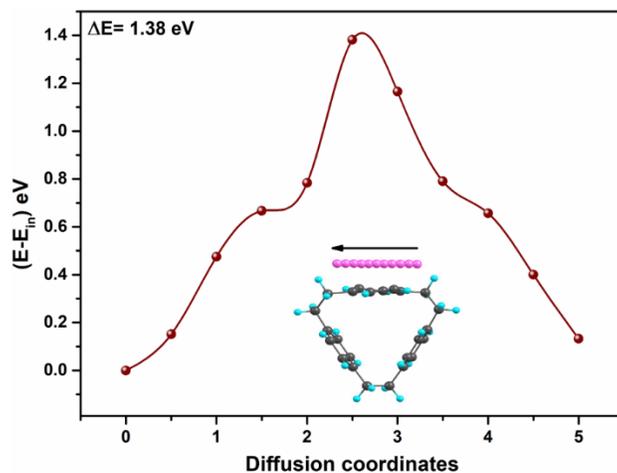

**Figure 2:** Diffusion energy barrier plot between energy difference and diffusion coordinates of Li atom on PCP222

The clustering of metal atoms on the substrate can reduce the hydrogen uptake capacity of the system as reported earlier [17]. The barrier of metal atoms diffusion energy ultimately decides whether or not the clustering will occur. With a small rise in temperature, if the Li atom migrated from its adsorption location, the possibility of metal-metal clustering would increase. Since, the binding energy of Li atom on the PCP222 is less than the cohesive energy of the isolated Li atom (1.63eV), we calculate if there is an energy barrier for diffusion of Li atom on PCP222 that can avoid the possibility of metal clustering. To calculate the energy barrier, we shift the Li atom over its adsorption site (on the benzene ring) by a small distance along the path shown in the Figure 2 and carried out the single point energy calculation. Then we exhibit the energy difference between initial and current step energy with the diffusion coordinate as illustrated in Figure 2. The figure shows presence of an energy barrier of 1.38 eV, that is sufficient to stop the Li atom from diffusing across the PCP222 and thus prevent the metal clustering. Therefore, our calculated energy barrier for diffusion of Li atom is high enough to prevent metal clustering over the studided PCP222 compound.

### 3.3 Interaction of $H_2$ with PCP222-Li

### 3.3.1 Adsorption Energy

**Table 1:** Average bond distance between carbon bridge (C-C), center of PCP222 benzene ring ($R_c$) and Lithium atom ($R_c$-Li), Lithium and hydrogen molecules (Li-$H_2$), and hydrogen Hydrogen (H-H) in Å. Average adsorption energy and successive desorption energy of PCP222-Li-$nH_2$ (n=1-5)

| Name of complex | Bridge C-C | $R_c$-Li | Li-H | H-H | $E_{ads}$ (eV) | $E_{des}$ (eV) |
|---|---|---|---|---|---|---|
| PCP222-Li | 1.542 | 1.735 | - | - | - | - |
| PCP222-Li-1$H_2$ | 1.542 | 1.745 | 2.124 | 0.753 | 0.171 | 0.171 |
| PCP222-Li-2$H_2$ | 1.541 | 1.742 | 2.083 | 0.757 | 0.159 | 0.147 |
| PCP222-Li-3$H_2$ | 1.541 | 1.767 | 2.159 | 0.753 | 0.148 | 0.127 |
| PCP222-Li-4$H_2$ | 1.541 | 1.811 | 2.243 | 0.752 | 0.134 | 0.089 |
| PCP222-Li-5$H_2$ | 1.541 | 1.813 | 2.478 | 0.751 | 0.113 | 0.030 |

To explore the storage capacity and characteristics of Li functionalized PCP222, we introduced the $H_2$ molecules in a sequential manner to PCP222-Li. Firstly we introduced a single $H_2$ molecule at around 2Å above the Li atom on PCP222 and allowed the structure to get relaxed. It is observed that, the $H_2$ molecule is adsorbed at a distance of 2.124 Å from the Li atom with an adsorption energy of 0.171 eV and the H-H bond length elongated by 0.01 Å. Sathe *et al.* studied the hydrogen storage capacity of Li functionalized PCP11 (PCP22) and reported the adsorption energy of first $H_2$ molecule ~0.13 eV (0.11 eV) [46, 45]. Our calculated adsorption energy is slightly higher, which is important in alkali metal doped $H_2$ storage material and leads to the increase in the desorption temperature. Further, we optimized the structures by adding $H_2$ molecules sequentially onto the PCP222-Li. On addition of second $H_2$ molecule to the system, the average $H_2$ adsorption energy calculated to be 0.159 eV/$H_2$. In this way, adsorption of 3rd, 4th and 5th $H_2$ molecules to

PCP222-Li, the average H$_2$ adsorption energy reduces to 0.148, 0.134 and 0.113 eV/H$_2$ respectively. When of more than five H$_2$ molecules are added to the system, they fly away from the Li atom and adsorption energy fall below 0.1 eV. We observed that the average adsorption energy decreases with increase in number of H$_2$ molecules in the system which is due the steric hindrance between the adsorbed H$_2$ crowed around the sorption centers and the increase in Li-H$_2$ distances (Table 1). The estimated data of adsorption energy and geometrical parameters of all the bare hydrogenated systems and presented in Table 1.

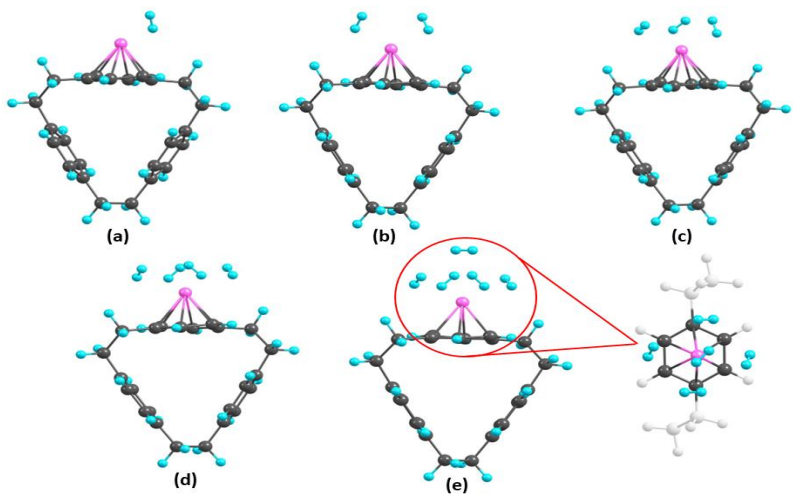

**Figure 3:** Optimized geometry of hydrogenated Li functionalized PCP222, (a) PCP222-Li-1H$_2$, (b) PCP222-Li-2H$_2$, (c) PCP222-Li-3H$_2$, (d) PCP222-Li-4H$_2$, (e) PCP222-Li-5H$_2$.

### 3.3.2 Electrostatics potential and Hirshfeld charges

To get a qualitative picture of electronic charge distribution over the surface of Li functionalized PCP222 and their hydrogen adsorbed systems during the hydrogen adsorbed, we generate and plotted the electrostatic potential map (ESP map) on the total electron density as depicted in Figure 4. The electronic charge distribution is used to identify the active adsorption site, where the hydrogen molecules can be introduced. The red and blue regions in the ESP plot reflects the aggregation and reduction of electronic charge density respectively. The variation in the charge density is plotted with the sequence of color code as red (highest electron density)> orange > yellow > green > blue (lowest electron density). The ESP map of PCP222-Li shows that the Li atom has the deficiency of electronic charges as marked by the dark blue region over the Li atom, this indicate that the Li atom is somewhat ionic and is prone to bind the guest H$_2$ molecules. When the first H$_2$ molecule added to the Li atom, the colour of the region over the Li changes from dark blue to light blue, demonstrating the charge transfer from C atom of PCP222 and adsorbed $H_2$ to the Li atom. Further sequential adsorption of $H_2$ molecules to PCP222-Li changes the colour of Li region from blue to light blue indicating additional charge transfer. The blue region over Li almost disappears on the adsorption of 5th $H_2$ molecules suggesting the saturation of hydrogen uptake and more guest $H_2$ are unlikely to be adsorbed. The exact charge transfer is determined by calculating the hirshfeld charges as discussed below.

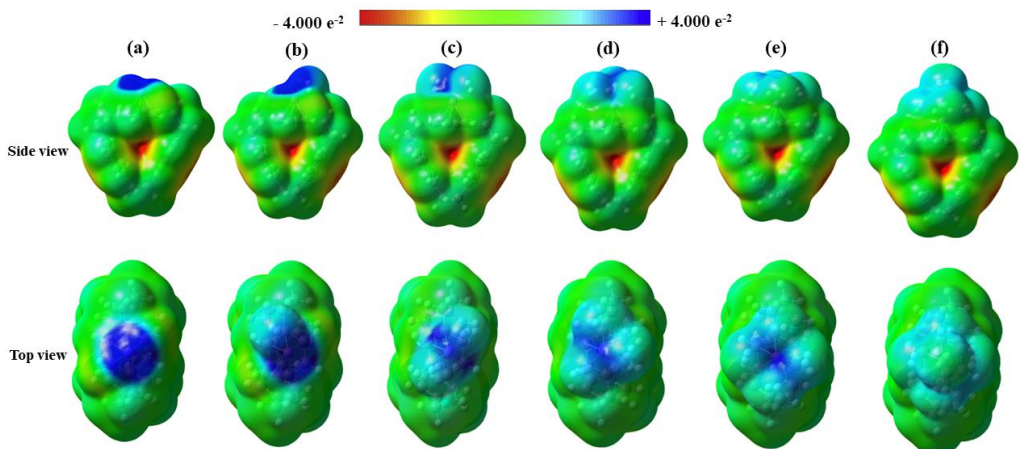

**Figure 4:** Electrostatics potential map of (a) PCP222-Li, (b) PCP222-Li-1$H_2$, (c) PCP222-Li-2$H_2$, (d) PCP222-Li-3$H_2$, (e) PCP222-Li-4$H_2$, (f) PCP222-Li-5$H_2$.

We have performed the hirshfeld charge analysis to quantify the charge transfer distributions on the Li functionalized PCP222 and their $H_2$ adsorbed systems. The computed average Hirshfeld charges on C atoms of benzene ring (Li functionalized site), Li atom, and adsorbed $H_2$ molecules with the number of hydrogen molecules is depicted in Figure 5. The average charges on C atom of benzene ring is noted to be -0.031 e which raises to -0.084 e with the functionalization of Li atom. The charge on Li atom of PCP222-Li is noted to be +0.354 e, which illustrate the transfer of charges from benzene ring to Li atom making the sorption center (Li) ionic and more suitable for $H_2$ adsorption. These results agree well with the aforesaid ESP analysis. On adsorption of the first $H_2$ molecule to PCP222-Li, the charge on C atom is reduced by 2.38% and at the same time the charge on Li atom is increased by 16.7 %. Further addition of hydrogen molecules follows the trend of decrease in charge on benzene ring and increase in charge on Li atom (Figure 5). These observations suggest that, the ionic Li atom polarize the guest $H_2$ molecules and the $H_2$ molecules are adsorbed to the sorption center via a charge polarization mechanism due to induced dipole developed in $H_2$ as suggested by the Neu-Rao-Jena [30]. It is noted that the electronic charge on

Li atom is raised by 41.36 % after the adsorption of the 5th H$_2$ molecule. The adsorbed H$_2$ molecules are found to have an average charge of 0.027e to 0.013 e.

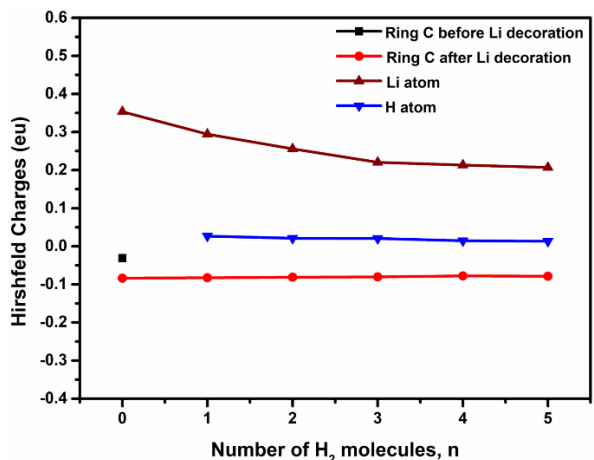

**Figure 5:** Hirshfeld charges before and after hydrogen adsorption on PCP222-Li

### 3.3.3 Bader's topological analysis and PDOS

The nature of interaction between the Li functionalized PCP222 and the adsorbed hydrogen molecules is analyzed using the topological Bader's quantum theory of atoms in molecules (QTAIM). The parameters of electron density distribution at the bond critical point (BCP), including the electron density ($\rho_{BCP}$), total electron energy density ($\mathcal{H}_{BCP}$), and Laplacian ($\nabla^2\rho_{BCP}$), are computed and given in Table S1 (in Supporting Information). The electron density (*r*) on C-C, and C-Li, of hydrogenated PCP222-Li estimated to be almost equal to that of bare host material, suggesting the post-adsorption chemical stability of the material. Additionally, the average $\rho_{BCP}$ values on H-H in PCP222-Li-5H$_2$ is 0.258 a.u which is same as that on isolated bare H$_2$ molecules (-0.263). This reveal that the adsorbed hydrogens are in molecular form during the adsorption. According to Kumar *et al.*, the positive value of $\nabla^2\rho_{BCP}$ indicated an electron density depletion in the region of bonding and implied a close-shell kind of interaction. We noticed there is no BCP between the Li and H atoms which implies no chemical bond between the Li atom and the adsorbed H$_2$ molecules and the interaction is purely closed-shell type resulting from the charge polarization as proposed by the Neu-Rao-Jena.

Figure 6 illustrate the density of state plot of Li and adsorbed H atoms of the hydrogenated PCP222-Li including the first and last (5th) H$_2$ molecules adsorbed on the system. When one hydrogen molecule is bound to the sorption center (Li), the *s*-orbital of the H$_2$ molecule appears below the Fermi level (E = 0) and stays unaffected as in the case of bare H$_2$ in Figure S2. This signifies that there is no hybridization between the Li and adsorbed $H_2$. This implies that the adsorption of $H_2$ molecule is owing to the induced dipole produced by charge polarization in $H_2$. With the adsorption of $5H_2$ molecules on PCP222-Li, the orbital of H atom splits into multiple peaks ranging from -16 eV to -4 eV. This implies that the adsorption weakens as the quantity of $H_2$ molecules increases in the host.

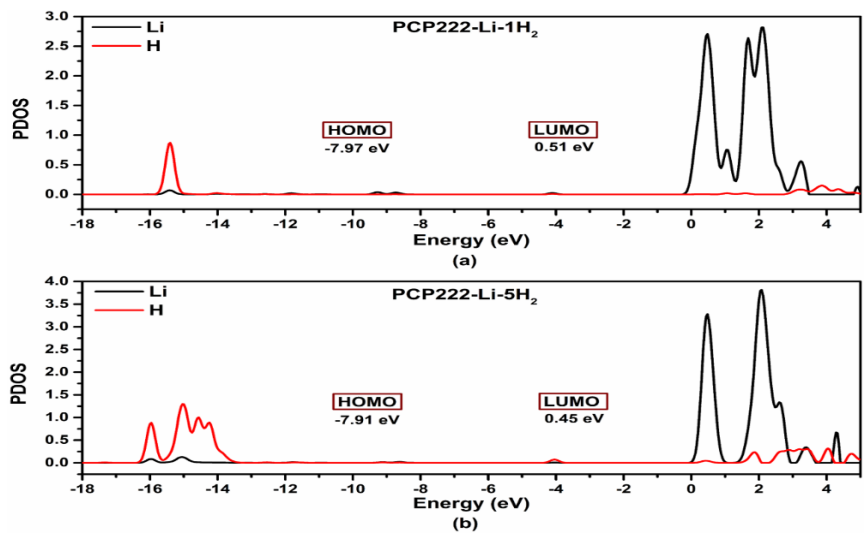

**Figure 6:** Partial density of state on Li and H atoms of PCP222-Li-$1H_2$ and PCP222-Li-$5H_2$

## 3.4 Thermodynamics and storage capacity

### 3.4.1 Storage Capacity

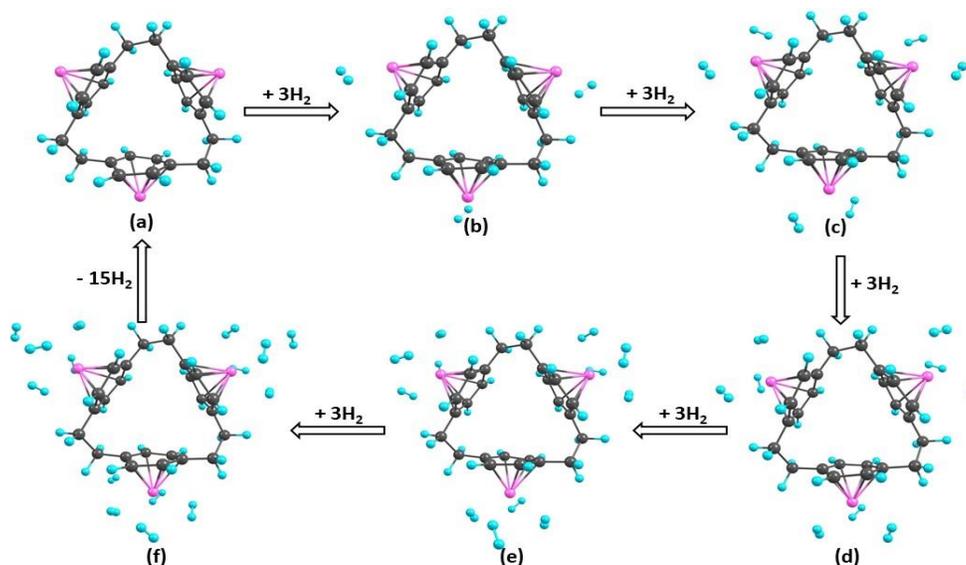

**Figure 7:** Optimized geometry of (a) PCP222-3Li, (b) PCP222-3Li-$3H_2$, (c) PCP222-3Li-$6H_2$, (d) PCP222-3Li-$9H_2$, (e) PCP222-3Li-$12H_2$, (f) PCP222-3Li-$15H_2$.

To investigate the optimum hydrogen storage capacity of the studied system, we functionalized the maximum possible number of Li atoms over each benzene ring of PCP222. The geometrical structure of three Li functionalized PCP222 ( PCP222-3Li) is shown in Figure 7 Further, we introduced $H_2$ molecules to each Li atom of PCP222-3Li sequentially as discussed in previous section (3.3.1). The computed average hydrogen adsorption energy and the geometrical parameters of all the hydrogenated systems are provided in the Table S2 (in Supporting Information). It is noticed that, the adsorption process of hydrogen molecules on PCP22-3Li is found similar to that of on PCP222-Li. On saturation of $H_2$ adsorption on PCP222-3Li, we found each Li atom can adsorb a maximum of $5H_2$ molecules resulting in total gravimetric density of 8.32 wt%. The estimated value of hydrogen storage capacity is fairly above the requirement of US-DOE for effective hydrogen storage systems. Our results can be compared with earlier reported $H_2$ gravimetric density on metal decorated carbon-based materials for hydrogen storage, for example, Li-decorated $C_{41}$ allotrope (7.12 wt%) [58], Li doped MOF impregnated with Li-coated fullerenes[59], Li-doped $B_4C_3$ monolayer (6.22 wt%) [4].

To develop a realistically usable hydrogen storage system, a significant quantity of hydrogen molecules must be adsorbed by the host material under achievable storage conditions. Further the adsorbed hydrogen molecules must also be efficiently desorbed at suitable temperature (T) and pressure (P). Thus, we estimated the quantity of adsorbed hydrogen that could be used at a accessible range of temperature (T) and pressure (P). To calculate the number of $H_2$ molecules remain adsorbed on PCP222-3Li (Occupation number) at different T and P, we calculated the empirical value of hydrogen gas chemical potential ($\mu$). Then the occupation number (*N*) is estimated by the following expression and plotted with various T and P in Figure 8(b)[60].

$$N = \frac{\sum_{n=0}^{N_{max}} n g_n e^{[n(\mu - E_{ads})/K_B T]}}{\sum_{n=0}^{n_{max}} g_n e^{[n(\mu - E_{ads})/K_B T]}} \qquad (5)$$

Here *N<sub>max</sub>* is the maximum number of $H_2$ molecules adsorbed at each Li atom on PCP222, *n* and *g<sub>n</sub>* represents the number of $H_2$ molecules adsorbed and configurational degeneracy for a *n* respectively. *k<sub>B</sub>* is the Boltzmann constant and -*E<sub>ads</sub>* indicates the average adsorption energy of $H_2$ molecules to PCP222-3Li. *m* is the empirical value of chemical potential of hydrogen gas at specific T and P, and is obtained by using the following expression [61].

$$\mu = H^0(T) - H^0(0) - TS^0(T) + K_B T \ln\left(\frac{P}{P_0}\right) \qquad (6)$$

Here $H^0(T)$, $S^0(T)$ are the enthalpy and entropy of H$_2$ at pressure $P_0$ (1 bar).

We can see in Figure 8(b) that the PCP222-3Li can adsorbed H$_2$ molecules giving rise to maximum hydrogen uptake capacity of ~8.32 wt% up to the temperature of 80 K and pressure of 30-60 bar. When the temperature rises beyond 80 K, the H$_2$ molecules begin to desorb from the PCP222-3Li and the gravimetric density closes to ~5.5 wt% (target of US-DOE by 2025) when the temperature reaches 180 K under the pressure of 30-60 bar. Further rise in temperature, the storage capacity of the PCP222-3Li fall below 4 wt% at 220 K and 40-bar. At a temperature range of 260 K-300 K and pressure range of 1-10 bar, the studied system shows a 100% desorption of hydrogen. Thus, we can propose the Li functionalized PCP222 as a low-temperature-adsorption and room-temperature-desorption hydrogen storage material. Under the room temperature (300 K), the studied system shows up to 8.32 wt % of usable hydrogen storage capacity with 100% reversibility. Thus, we believe that, our studied material Li functionalized PCP222 can be used as an efficient hydrogen storage material satisfying the criteria of US-DOE.

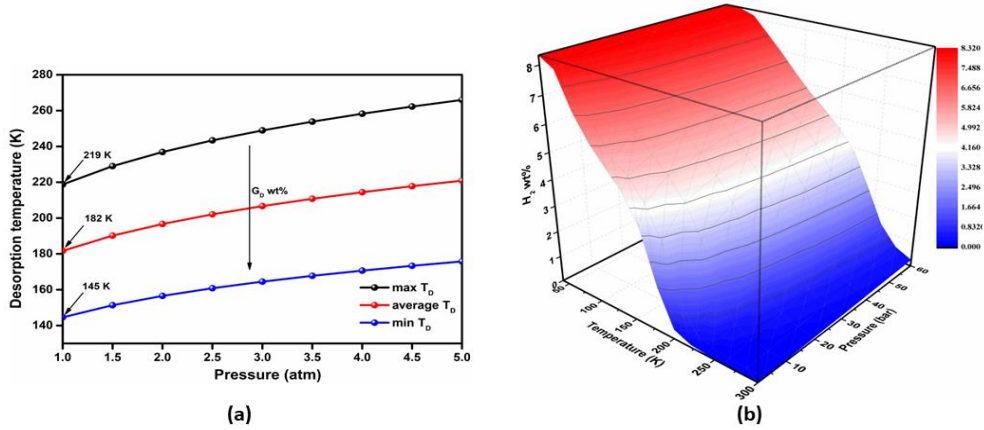

**Figure 8:** Plot of Van't-Hoff desorption temperature for Li functionalized PCP222 at different temperature and pressure.

### 3.4.2 Desorption temperature

For a reversible hydrogen storage media, it is crucial to estimate the desorption temperature of hydrogen molecules. We have estimated the desorption temperature ($T_D$) of H$_2$ for the Li functionalized PCP222 using the Van't Hoff equation [17].

$$T_D = \left(\frac{E_{ads}}{K_B}\right)\left(\frac{\Delta S}{R} - \ln p\right)^{-1} \quad (7)$$

Where, $E_{ads}$ represents the computed hydrogen adsorption energy, $K_B$, and R denotes for the Boltzmann constant and R the gas constant respectively. P represent s the equilibrium pressure (we take a range of 1 to 5 atm with an increment of 0.5 atm) and $\Delta S$ is the entropy change of hydrogen from its gaseous state to liquid state [62]. Using the highest and lowest adsorption energy of system (with minimum and maximum H₂ gravimetric density, respectively), the maximum and minimum desorption temperatures ($T_{Dmax}/T_{Dmin}$) are determined. While, $T_{Dmin}$ denotes the minimum temperature necessary to initiate the desorption of H₂ molecules, the $T_{Dmax}$ is the temperature required for complete desorption process. The estimated desorption temperatures along with the equilibrium pressure is depicted in Figure 8(a). The minimum and maximum temperatures for H₂ desorption are determined to be 145 K and 219 K, respectively, at 1 atm pressure. The estimated average $T_D$ of Li functionalized PCP222 is 182 K at 1 atm. This result reveals that, the system can adsorb its full capacity H₂ at cryogenic temperature and desorb all the H₂ molecules bellow room temperature at 1 atm pressure. However, the desorption temperature can be increases by increase in the equilibrium pressure as presented in Figure 8(a) and as discussed above.

## 3.5 Molecular dynamics simulations

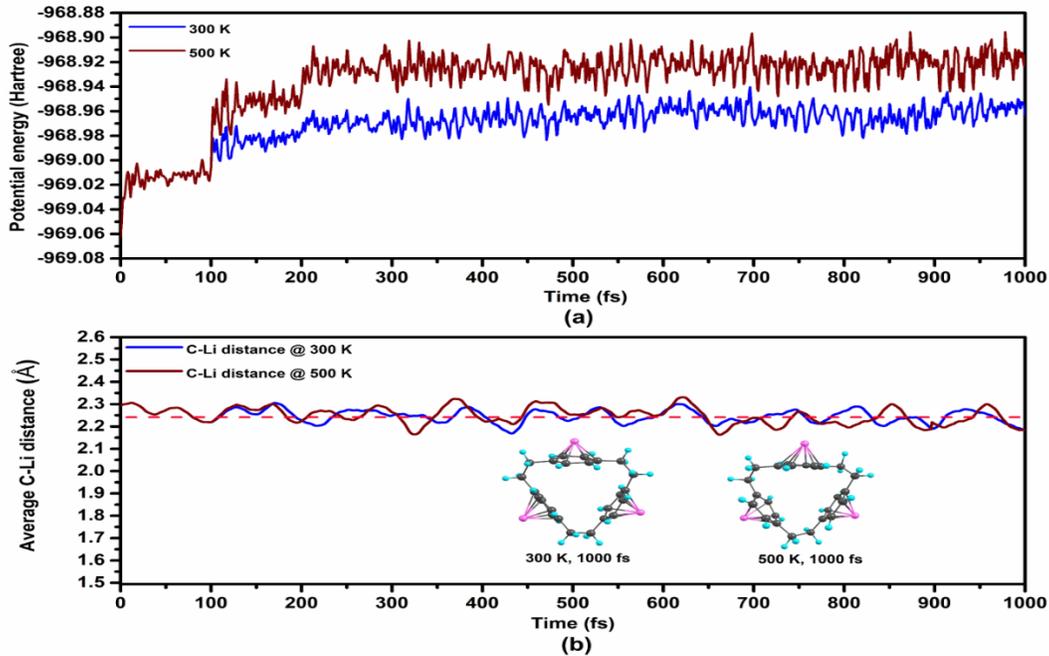

**Figure 9:** (a) Potential energy trajectories of hydrogenated PCP222-3Li and (b) Time evolution trajectory of average bond length between the Li atom and C atoms of PCP222 at 300K and 500K,

To validate the reversibility of hydrogen molecules on PCP222-3Li estimated by the DFT computation, we have carried out molecular dynamics (MD) simulations using the atomistic density matrix propagation (ADMP). ADMP is an extended Lagrangian approach to MD, that uses the gaussian basis function and propagates the density matrix. The ADMP-MD simulations is performed on system with highest storage capacity (PCP222-3Li-15$H_2$), at two different temperatures of 300K and 500 K for total time of 1 ps with the time step of 1fs. During the simulations the temperature (kinetic energy thermostat) is maintained by the velocity scaling approach and at every 10 fs, time step, the temperature is checked and corrected. The time evolution potential energy trajectories and the snapshots are depicted in Figure 9(a) and Figure S3 (in supporting Information) respectively. The MD simulations at 300 K and 1ps illustrate that almost all the $H_2$ molecules fly away from the sorption centers, except 1$H_2$ at each center. Simulations at 500 K shows that all the $H_2$ molecules are desorbed from the host material keeping the host structure intact. This result suggests that the hydrogen storage in Li functionalized PCP222 is reversible in process.

For a viable reversible hydrogen storage material, it is important that the host material must not distorted above the hydrogen desorption temperature. To investigate the solidity of host material (PCP222-3Li), we performed the MD simulations on the bare host structure (PCP222-3Li) at room temperature (300 K) and considerably above the $H_2$ desorption temperature (500K) using ADMP. The molecular dynamics simulations are performed for 1 ps with a time step of 1 fs. The time evolution trajectory of average distance between Li atom and the carbon atoms of PCP222 benzene rings is plotted in Figure 9(b). We noticed that the PCP222-3Li structure stays stable at 500 K and almost no change in C-C and C-H bond distance is observed. The trajectory of average bond length between the Li atom and C atoms of PCP222 benzene rings seem oscillate but the mean value (2.25 Å) and the variation is minimal. This validates the structural integrity of the host material above the $H_2$ desorption temperature. Moreover no Li clustering is also noticed after desorption as discussed earlier in Section 3.2.1. Thus, we believe that PCP222-3Li can be considered for feasible reversible hydrogen storage material.

## 4 Summery and Conclusion

In this study, we investigated the thermodynamical stability and hydrogen storage capacity of Li functionalized [2,2,2]paracyclophane, using the density functional theory. The Li atoms are found to bind with the PCP222 via Dewar mechanism and no clustering of Li atoms over PCP222 was noticed. Each Li atom on PCP222 could adsorb up to 5$H_2$ molecules via charge polarization mechanism with an average $H_2$ adsorption energy in the range of 0.12 - 0.17 eV/$H_2$, indicating physisorption type of adsorption. Moreover, the average H-H bond distance got elongated by 0.01 Å, during the adsorption process, which implied that the adsorbed $H_2$ were in molecular form and this fact was also confirmed by the charge distribution analysis. When three Li atoms were functionalized on PCP222, the $H_2$ gravimetric capacity of the system was up to 8.32 wt% which was fairly above the US-DOE requirements for practical hydrogen applications. During saturation of $H_2$ adsorption, the host material displayed no significant change in geometry. The thermodynamic usable hydrogen capacity was found up to ~8.32 wt% at the temperature of 80 K and pressure of 30-60 bar. On further increase in temperature, up to 180 K under the pressure of 30-60 bar, the PCP222-3Li hydrogen uptake capacity approached 5.5wt% which is the target of DOE by 2025. At a temperature range of 260 K-300 K and pressure range of 1-10 bar, the PCP222-3Li system showed 100% desorption of $H_2$. Molecular dynamic simulation confirmed that at 300 K, almost all the $H_2$ molecules flied away except 1$H_2$ at each center. Simulations at 500 K showed that all the $H_2$ molecules are desorbed from the host material keeping the structure of the host structure intact. Since, there is no experimental works reported on Li functionalized PCP222 for hydrogen storage, we hope our computational work will contribute significantly to the research of hydrogen storage in macrocyclic compounds and provide supporting reference for the future experiments.